# Stimulus and correlation matching measurement technique in computer based characterization testing

A. M. Dorman, test engineer, consulting, Israel

**Abstract**. Constructive theory of characterization test is considered in a proposed article. The theory is applicable to an electronic nano devices characterization. E.g. is a current-voltage test particularly differential conductance test. Another application is an auger current-electron energy spectroscopy test. Generally small response of device under test on an applied stimulus is masked by an unknown deterministic background and a random noise. Characterization test in this signal corruption scenario should be based on correlation measurement technique of device response on applied appropriate i.e. optimal stimulus with optimal reference signal. The problem of co-synthesis of stimulus and the reference signal is solved here. This theory provides constructive solution to improve measurement accuracy and speed of characterization test. Proposed test system implementation, for example, may improve measurement accuracy or speed of auger current test by more than 100 times.

**Introduction**. The major challenge in small scale device characterization is dependence plot and its informative parameters measurement. The dependence may be current - voltage curve or auger current - energy of electrons curve.
Nano devices electrical characteristics are affected by quantum behavior, Figure 1, [1].

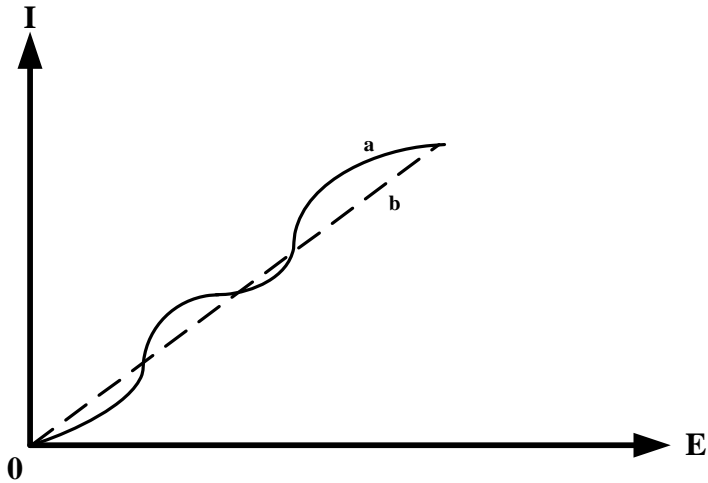

Figure 1 Nano device current - voltage characterization
a - quantum affected dependence
b - ohm low obeyed dependence

Thus, measurement of a current-voltage curve differential conductance $dI(E)/dE$ or a current-voltage $I(E)$ curve **a** divergence from ohm law straight line **b** as it shown on Figure 1 is very important in nano devices characterization [1]. Informative parameter is the position $E_p$ of auger peak on the electron energy scale $E$, full auger current $I_c$ i.e. peak **a** area over a background **b** dashed line in auger electron spectroscopy test, Figure 2. Auger peak current value $I_c(E)$ for specified energy value $E$ should be also considered as an informative parameter, Figure 2.

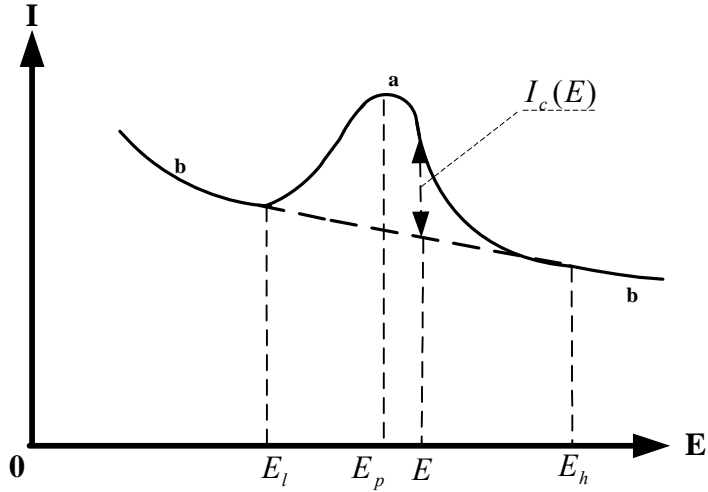

Figure 2 Auger current - electron energy characterization
a - auger current peak
b - secondary electrons background

It is assumed that auger current $I_c(E)$ is a difference between an acquired current-electron energy distribution i.e. curve **a** and background $I_b(E)$ i.e. secondary etc. electrons current-energy distribution curve **b** as it shown on Figure 2.

Full auger current with this assumption is $I_c = \int_{E_l}^{E_h} I_c(E)dE$ , where $E_l$ and $E_h$ are energy scale low and high boundaries of auger peak correspondingly, Figure 2. Quality metrics of the test i.e. its accuracy and speed is limited by the influence of a non-informative masking unknown background $I_b(E)$ and by different components of a random electrical noise $\xi(E,t)$, where $t$ is a time. These components are thermal noise, shot noise etc.. Testing must be non destructive. This requirement implies that low level stimulus is used. As a result of this, random noise becomes one of a prevalent cause of measurement errors and the factor which reduces test speed. The second factor, which reduces accuracy and test speed, is non-informative background interference. For example, linear current-voltage component of acquired curve obeyed ohm law, may be considered as background in case of nano device characterization. This linear component is masking small non-linear informative portion of acquired signal. Secondary electron current energy distribution is a masking background in case of auger electron spectroscopy.

The problem of test quality metrics improvement has respect to a variety of applications. This actual issue is the subject of numerous articles. Nevertheless the optimum test methods guaranteeing extreme quality i.e. accuracy and speed of the test are not completely examined.

**Control, processing, measurement.** Informative parameter $\theta$ is a parameter of informative signal $I_c(E)$, generated by the device under test (DUT). Informative parameter may be, for example, an area under auger current peak or a current-voltage curve deviation from ohm law for a specified values of voltage stimulus $E$. Test system input signal $I(E,t)$ is a mixture of DUT output informative signal, i.e. characteristic $I_c(E)$, with background $I_b(E)$ and noise $\xi(E,t)$, as follows $I(E,t) = I_c(E) + I_b(E) + \xi(E,t)$. Main function of optimal test system is to estimate informative parameter by the best way: for example minimum errors for fixed speed of testing. Ideally test system suppresses random noise so that informative parameter should be recovered from the curve $I(E,t) = I_c(E) + I_b(E)$.

Informative parameters may be separated into two groups. The first group parameters, like auger electrons peak location on the energy scale, may be named a *non energy dependant parameter*, because peak amplitude stretch operation, i.e. signal energy variation, has no impact on this parameter value. The second group parameters, like full auger electrons current, may be named an *energy dependant* ones. This means that peak amplitude stretch operation will modify the value of informative parameter.

*Non energy dependent* parameter measurement should be generally based on multistep sequential statistical non-linear estimation. It can be shown, that under very general assumptions informative optimal *energy dependent* parameter estimation should be one step linear statistical estimation. *Energy dependent* parameter $\hat{\theta}$ estimation, related to energy $E_c$ can be done as follows,

$$\hat{\theta}(E_c) = F\{\int_E I(E,t) K_w(E,E_c) dE\}, \tag{1}$$

where $F(.)$ is a low frequency filtration operator. Integration is performed in the range of variation of stimulus $E$. The limits of integration may be as follows, $\{E_c + E_m/2, E_c - E_m/2\}$. $E_m$ is the range of variation of the stimulus used for parameter $\hat{\theta}(E_c)$ estimation. Parameter $\hat{\theta}(E_c)$ in (1) may be, for example, auger current measurement i.e. estimation result. This result is related to the value $E_c$ of auger electrons energy, for which the DUT characteristic is to be tested. The weighting function $K_w(E,E_c)$ in (1) should be carefully designed to minimize background dependent component of error. Background suppression requirement is an initial specification for weighting function choice. Ideally background should be rejected

$$\int_E I_b(E) K_w(E,E_c) dE = 0.$$

Specification like an example above should be considered as the weighting function $K_w(E,E_c)$ constraint, when weighting function is varied to minimize a random error. Let's suppose, for example, that measured parameter is a current of selected electrons, whose energies are within low $E_l$ and high $E_h$ limits. Assume also that background is close to a linear function of $E$. In this case, the following weighting function, for example, will provide the background suppression:

$K_w(E, E_c) = 1$, when $E \in [E_l, E_h]$;

$K_w(E, E_c) = 0.5(E_l - E_h)[\delta(E - E_l) + \delta(E - E_h)]$, when $E \notin [E_l, E_h]$.

$\delta(E)$ is a delta function.

If a curve divergence from a background is under test, e.g. an auger peak current for electron energy $E_c$ is a parameter to be measured in the test, the weighting function may be represented as follows

$K_w(E, E_c) = \delta(E - E_c) - f(E, E_c)$.

Function $f(E)$ must be chosen in such a way as to suppress the background [2]. In particular this function can be specified in parametric form. For example $f(E, E_c) = \sum_i a_i(E_c)\delta(E - E_i)$.

These coefficients $a_i(E_c)$ must be chosen so as to minimize the impact of a background on the measurement result. Specifically Dolph-Chebyshev coefficients may be used in this equation.

There are some ways of method (1) implementation. Practically it can be implemented as a meter of signal correlation. DUT output signal $I(E,t)$ in (1) caused by the applied stimulus $E(t)$ is to be correlated with the reference signal $u(t)$ of meter as follows

$$\hat{\theta} = \int_t I[E(t), t] \times u(t) dt .\qquad(2)$$

Reference signal $u(t)$ must be chosen so that the correlation estimate (2) i.e. parameter $\hat{\theta}$ would coincide with the definition (1).

Stimulus i.e. scanning signal $E(t)$ in (2) consists generally of a constant or a slowly varying control $E_c(t)$ and a fast modulation $E_M(t)$ component

$$E(t) = E_c(t) + E_M(t).$$

Estimated parameter is related to a slow control component $E_c(t)$. This component controls a weighting function and relates to the shift of the weighting function along the axis $E$, $K_w(E, E_c) = K_w(E - E_c)$. Specifying $E_c$ in the formulas is not critical and will be omitted further.

It should be emphasized that the scanning signal $E(t)$ may only contain a constant control $E_c(t) = const$ and fast modulating components $E_M(t)$. In this case, the direct control of the weighting function should be used to estimate informative parameters related to the different points of the scale $E$. Modulation technique allows to measure an informative parameter in a frequency band in which the interfering noise is small. Modulating signal $E_M(t)$ is a periodical signal $-E_m/2 \leq E_M(t) \leq E_m/2$ with a period $T$ and range $E_m$. A control signal $E_c(t)$ should be varied to choose different areas of curve under test i.e. current-voltage, current-energy of electrons etc. curves. The rate of change of a control signal $E_c(t)$ must be limited as follows to avoid aliasing errors

$$|dE_c(t)/dt| \leq \pi/T\omega_B .$$

Parameter $\omega_B$ [radian/volts] in this equation is the upper cut-off frequency of dependence i.e. spectrum of a function $I(E)$ in a frequency domain related to the stimulus scale $E$.

If this condition is met, modulating signal and reference signal are related as it follows

from equations (1) and (2) by the following differential equation

$$T^{-1} \sum_{i=1}^{V(E)} u_{eq}[t_i(E)] / |E'_M|_{t=t_i} - K_w(E) = 0, \quad (3)$$

where

$$u_{eq}(t) = \sum_{n=-\infty}^{\infty} \overset{*}{S}(n\omega_0) \overset{*}{U}(n) \exp(jn\omega_0 t). \quad (4)$$

$u_{eq}(t)$ may be named equivalent reference signal because equation (4) takes into account not only the shape of physical reference signal $u(t)$ of correlation meter, but also the transfer function of the DUT output signal transducer. Specifically $\overset{*}{S}(\omega)$ is the transfer function of the DUT output signal transducer, $\overset{*}{U}(n)$ is the complex spectrum component of a physical reference signal $u(t)$ at a frequency $n\omega_0$, $\omega_0 = 2\pi/T$ is the modulating signal frequency.

Time moments $t_i$ in (3) can be found from the following equation $E_M(t_i) = E$, $i = \overline{1, V(E)}$, where $V(E)$ is the number of solutions $t_i$.

Let's suppose that weighting function has been chosen in such a way so that to provide required background suppression. Now we need to minimize the impact of a random noise on the accuracy of the test. It means that we must find such signals, i.e. modulating $E_M(t)$ and reference $u(t)$, that minimize the variance of the random error. Random error variance

is $D_n = \sum_{l=1}^{\infty} P_{nl} |\overset{\bullet}{U}(l)|^2$, (5)

where $P_{nl}$ is the noise power in the signal band at a frequency $l\omega_0$.

Constraint (3) on the shape of both modulating $E_M(t)$ and reference $u(t)$ signals should be taken into account when variance (5) is minimized. Let's suppose that $\overset{*}{S}(n\omega_0) = const(n)$, $P_{nl} = P_n$ and modulating signal is a continuous function. In this case, it can be shown [2] that the optimal reference signal $u(t)$ of a correlation measuring channel is bi-level

$$u(t) \equiv u_{eq}(t) \equiv sign\{K_w[E_M(t)]\}, \quad (6)$$

and optimum modulating signal $E_M(t)$ is

$$\Gamma_w[E_M(t)] = T^{-1}\Gamma_w(E_M/2) \times (t + T/2), \quad (7)$$

where $\Gamma_w(E) = \int_{-E_m/2}^{E} |K_w(x)| dx$.

If the weighting function is given in discrete form,

$$K_w(E) = \sum_{i=1}^{M} W(E_i)\delta(E - E_i), \quad (8)$$

it means, that the modulating signal must have a stepwise shape. $M$ is the number of individual values of a modulating signal. In this case, optimal modulating signal is a sequence of a discrete values $E_i$ for the corresponding time intervals $\tau_i$. I.e. modulating signal steps are the constants

$$E_M(t) = E_i, \; i = \overline{1, M}$$

during the time intervals $\tau_i$,

$$\tau_i = T|W(E_i)| / \sum_{n=1}^{M} |W(E_n)|,$$

within the period of modulation $T$.

Optimal reference signal is bi-level

$$u(t) \equiv sign[W(E_i)]$$

as in example above.

Optimum variance of a random error will be

$$D_n = P_n \times \Gamma_w^2(E_m/2) \quad . \tag{9}$$

It is worth noting that the weighting function can be optimized in turn in order to minimize optimum variance (9). Background suppression specification, informative characteristics and parameters distortion specification etc. requirements should be considered as constraints on weighting function in this optimization problem.

**Simultaneous measurement and testing** of dependence $I_c(E)$ and its derivative $dI_c(E)/dE$ may be especially useful for characterization of the DUT. In this case, test system has two correlation meter channels. One for dependence $I_c(E)$ and second for derivative $dI_c(E)/dE$ curve testing. Let's suppose that a modulating signal is discrete i.e. $E_M(t) = E_i$. Then the optimum solution is as follows.

Optimum modulating and reference signals of both channels have a stepwise shape and are constants during time intervals $\tau_i = \dfrac{T\sqrt{\mu_c W_c^2(E_i) + \mu_d W_d^2(E_i)}}{\sum_{i=1}^{M}\sqrt{\mu_c W_c^2(E_i) + \mu_d W_d^2(E_i)}}$ , $i = \overline{1, M}$ ,

within the period of modulation $T$. Reference signal for $I_c(E)$ testing

$$u_c(t) \equiv W_c(E_i)/\tau_i$$

is not bi-level in contrast to the previous case.

Reference signal for derivative $dI_c(E)/dE$ testing is

$$u_d(t) \equiv W_d(E_i)/\tau_i.$$

It is also not a bi-level signal.

$W_c(E_i)$ and $W_d(E_i)$ in these equations are coefficients of a discrete weighting functions, similar to formula (8) above, targeted to estimate respectively dependence $I_c(E)$ and its derivative $dI_c(E)/dE$.

Parameters $\mu_c$ and $\mu_d$, where $\mu_c \geq 0$, $\mu_d \geq 0$, $\mu_c + \mu_d = 1$, affect the variance of the random errors in both channels of a correlation meter. By varying the ratio of these coefficients, we can establish the necessary balance between the random errors in both channels.

**Colour noise**. Let's suppose that an informative signal is accompanied by a "coloured" noise i.e. $\overset{*}{S}(n\omega_0) \neq const(n)$ and/or $P_{nl} \neq const(l)$. Typically power density of a noise has a well-expressed minimum at some frequency. A good approach to the optimum solution in this case can be based on the use of a narrow-band reference signals. Ideally, the reference signal has to be harmonic $u_{eq}(t) \equiv u_0 \cos\omega_0 t$, and its frequency should be equal to the frequency where the spectral density of noise is a minimum. Of course, modulation frequency must be chosen so that the informative signal spectrum would be concentrated in the frequency domain area with minimal harmful interference. Modulating signal should be varied to minimize random measurement error variance $D_n$. If weighting function meets the condition listed below, a random error variance is subject, as it follows from (3) and (4), to a simple formula

$$D_n = P_n \Gamma_{TP}^2(E_m/2) * \pi^2/8. \tag{10}$$

Variance of the random error (10) is only $\pi^2/8$ times larger then the optimum value (9). The condition mentioned above is $sign\{K_w[E_M(t)]\} = sign[u_{eq}(t)]$.

Let's suppose that weighting function $K_w[E_M(t)]$ and the reference signal retain their sign during the time $\tau_l$, $l = 1,2,...$, equal to a multiple of the reference signal half-period. I.e. the time $\tau_l$ corresponds to the $E_M(t) \in [E_{l-1}, E_l]$, whereas the $K_w[E_M(t)]$ keeps its sign constancy. This condition constrains the choice of the weighting function and the modulating signal frequency, as follows $\tau_l = (2\pi/\omega_0)[\int_{E_{l-1}}^{E_l} |K_w(E)| dE / \Gamma_w(E_M/2)]$. Generally period of the optimal modulating signal is longer than the period of the reference signal.

**Harmonic modulation and lock-in detection**. Harmonic stimulus in combination with lock-in detection technique is conventionally used to acquire derivative $dI_c(E)/dE$ of current-voltage, current-energy of auger electron etc. characteristics. The implementation of a measurement channel can be performed as a lock-in detector or a correlation meter with harmonic reference signal. The output signal of the measurement channel differs from an ideal derivative $dI_c(E)/dE$ of a characteristic dependence. The more amplitude of the modulating signal, the more output signal of the measurement channel differs from an ideal curve $dI_c(E)/dE$. Therefore, output signal of the measurement channel should be corrected. Output signal of the lock-in detector is close to the derivative of the tested curve in case of small amplitude of the modulating signal. In this ideal situation output signal of the lock-in detector must be integrated to restore the shape of curve, e.g. auger peak, and integrated once more to obtain an area of the curve above the background, e.g. full auger current. A more sophisticated linear correcting filtration is required in a general case of arbitrary amplitude of the modulating signal. Correction reduces systematic errors but increases random one.

Optimal methods provide a smaller random error than a conventional method based on the lock-in detection. The gain in reducing the variance of the random error depends on the level of systematic error.

Let's suppose that two test systems are compared. The first one is the optimal system. The second is a conventional system with lock-in detector and integrators. Both systems have equal systematic errors and test speed. If a signal excess area above a background i.e. full auger current $I_c$ is under test, and systematic errors is about 3-5 %, optimal system provides up to 100 times smaller variance of a random error than the conventional system has. In the case of a curve $I_c(E)$ and its derivative $dI_c(E)/dE$ characterization testing, optimal system under the same conditions has respectively 20 and 6 times less variance of the random error than the conventional system has [2]. Speed of testing in the optimal system exceeds the rate of testing in the conventional system in the same proportion under the assumption that the both systems have the same systematic and random errors.

**Multi-dimension characterization**. In some tests, the object is characterized by the spatial distribution of the signal or shortly - by spatial map $I_c(x,y)$. It may be the dependence of electron emission current related to spatial coordinates of the DUT surface point $(x,y)$, where the emission is excited. For example, auger current map $I_c(x,y)$ characterizes the distribution of chemical admixtures over the surface of the DUT. Just as in the case of one-dimensional characteristic tests, as mentioned above, the effect of background and random noise is the principal obstacle also in this situation. Random noise as well as unknown background signal $I_b(x,y)$ e.g. current of secondary etc. electrons should be suppressed.

The primary electron beam is scanning over the surface area of the DUT, $x = x_M(t)$, $y = y_M(t)$. DUT output signal passes over correlation meter. The measuring value is defined by (1), but the weighting function in this equation must be replaced by the following two-dimensional function $K_w(x,y)$. Systematic errors are referred to the properties of a weighting function $K_w(x,y)$ and associated with incomplete background suppression and distortion of informative characteristic. Let's assume that the weighting function is defined at the points $(i,j), i = 0,1,2,...; j = 0,1,2,...$ of the tested surface with coordinates $x_i, y_j$, as follows $K_w(x,y) = K_w(x_i, y_j)$. Optimal $I_c(x,y)$ mapping is based on the correlation measurement [2]. Like the one-dimensional case considered above, the optimal reference signal of the correlation measurement channel is a bi-level signal $u(t) \equiv sign\{K_w[x_M(t), y_M(t)]\}$. Two components of the scanning vector $x_M(t) = x_i$, $y_M(t) = y_i$ are constant values during time intervals

$$\tau_{ij} = T \,|\, K_w(x_i, y_j) \,|\, / \Gamma_w. \tag{11}$$

Scan period T is equal to $\sum_i \sum_j \tau_{ij}$. $\Gamma_w$ definition is similar to a one-dimensional definition discussed above.

This theory may be adapted to the testing of "dynamic" DUT. As an example, let's assume that response of the DUT depends on applied stimulus $E_M(t)$ and its time derivative $dE_M(t)/dt$. Stimulus in this case is a two-dimensional vector $\{x_M(t), y_M(t)\}$ whose components are defined as follows $x_M(t) = E_M(t)$, $y_M(t) = dE_M(t)/dt$.

Stimulus performs scanning in the area $A$, $\{x_M(t), y_M(t)\} \in A$, where characterization testing planned to be done. The trajectory can be chosen, for example, as the scanning raster, consisting of a segments with a constant scanning speed. Further we assume that weighting function is specified for the set of a scanning speed samples as follows: $K_w(x,y) = K_w[E_M(t), dE_M(t)/dt] = K_w(E_i, E_j')$. $(E_i, E_j')$ are the coordinates of the points in the scan area $A$ for which the weighting function is defined. Thus, taking into account these assumptions, the optimal solution follows from the equation (11). Trajectory of the scanning stimulus vector $(E_M(t), dE_M(t)/dt)$ passes through all the points $(E_i, E_j')$ of the scan area, where the weighting function is specified, and consists of an ascending and descending branches.

There is a freedom in choosing a trajectory path under the mentioned above condition (11). Anyway, the branches of the trajectory with opposite directions must have equal total length. This condition may be named "balanced packing". Let's suppose that a trajectory is composed of a number linearly increasing and decreasing branches. Then a "balanced packing" condition means, taking into an account the equation (11), that

$$\sum_i \sum_{E_j'>0} |E_j'| \times |K_{TP}(E_i, E_j')| = \sum_i \sum_{E_j'<0} |E_j'| \times |K_{TP}(E_i, E_j')|.$$

If the weighting function is given as a continuous function of the coordinate $x$, summation over the corresponding index $i$ should be replaced by the integration over this coordinate $x$.

**Considerations for implementation**. Figure 3 shows one of a possible software and/or hardware implementation of the test system.

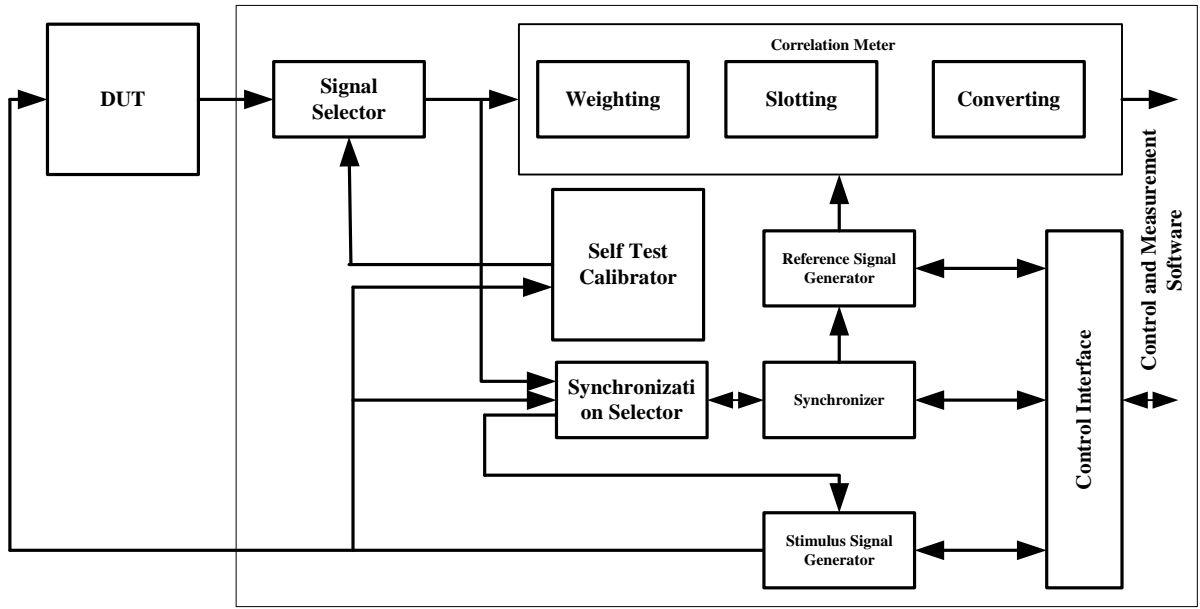

Figure 3. Voltage-current dependence test system.

DUT response on a stimulus is processed and measured by the measurement channel named *Correlation Meter*. *Stimulus Signal Generator* and *Reference Signal Generator* are synchronized. Reference signal may be synchronized by a stand alone clock generator located in *Synchronizer* or by a signal of *Stimulus Signal Generator* or by the DUT response via *Synchronization Selector*. DUT response is weighted and integrated in a time slots and after that converted into digital form. *Correlation Meter* also may work in a **slot measurement mode**, when output signal is averaged in the time slot intervals. Slot intervals are synchronized with stimulus signal in all modes. Final correlation may be calculated in software. A **slot measurement mode** may be useful if reference is a bi-level signal, as shown above. *Control and Measurement Software* communicates with all three channels i.e. C*orrelation Meter*, S*timulus Signal Generator* and S*ynchronizer* via test system C*ontrol Interface* module. Some important components of the test system like preamplifiers and so on are not shown on Figure 3.

Test system provides harmonic i.e. conventional modulating mode and also discussed above optimal modulating modes for $dI_c(E)/dE$, $I_c(E), I_c$ characterization tests. Simultaneously measured dependence $I_c(E)$ and its derivative $dI_c(E)/dE$ test mode can be also supported. Measurement result is related to a specified value $E = E_c = E_c(t)$ of stimulus scale $E$. Weighting function can be selected and configured in a self-test calibration mode. Effectiveness of a selected weighting function may be verified. Output signal of a real DUT is replaced in self-test calibration mode by the output signal of *Self Test Calibrator* module via *Signal Selector*. Self-test calibration mode especially convenient for arbitrary weighting function verification before its practical usage in an arbitrary measurement tests.

**Conclusions**. The problem of optimal measurement and stimulus control for characterization testing i.e. current - voltage, auger current - electron energy and so on is formulated and solved in this article. It is shown that the optimal measurement should be based on the correlation techniques. Optimal control of a stimulus and an optimal reference signals of the correlation measurement channels are proposed for the main modes of operation. The theory is applicable to an electronic nano devices characterization tests, auger electron spectroscopy tests. Proposed solutions may significantly improve performance i.e. accuracy and speed of characterization test modes.